\documentclass[useAMS,usegraphicx,usenatbib]{mn2e} 
\usepackage{graphicx}
\usepackage{times} 
\usepackage{amssymb}
\usepackage{amsmath}
\usepackage{lscape}
\usepackage{url}
\newif\ifAMStwofonts
\AMStwofontstrue

\def\chandra{{\it Chandra}}
\def\hst{{\it HST}}

\def\xmm{{\it XMM-Newton}}

\def\ciao{\hbox{\rm CIAO}}
\def\ciaov{\hbox{\rm CIAO\thinspace v3.2.2}}
\def\caldb{\hbox{\rm CALDB\thinspace v3.1.0}}

\def\fri{\hbox{\rm FR\thinspace I}}
\def\frii{\hbox{\rm FR\thinspace II}}

\def\s{\hbox{$\rm\thinspace s$}}

\def\yr{\hbox{$\rm\thinspace yr$}}

\def\hz{\hbox{$\rm\thinspace Hz$}}

\def\ghz{\hbox{$\rm\thinspace GHz$}}

\def\a{\hbox{$\rm\thinspace\AA$}}

\def\kpc{\hbox{$\rm\thinspace kpc$}}
\def\mpc{\hbox{$\rm\thinspace Mpc$}}

\def\as{\hbox{$\rm\thinspace arcsec$}}
\def\am{\hbox{$\rm\thinspace arcmin$}}

\def\pcmsq{\hbox{$\rm\thinspace cm^{-2}$}}
\def\kmpspmpc{\hbox{$\rm\thinspace km~s^{-1}~Mpc^{-1}$}}

\def\kev{\hbox{$\rm\thinspace keV$}}

\def\mjy{\hbox{$\rm\thinspace mJy$}}

\def\jy{\hbox{$\rm\thinspace Jy$}}
\def\jypb{\hbox{$\rm\thinspace Jy/beam$}}

\def\cts{\hbox{$\rm\thinspace counts$}}

\def\photpkevpcmsqps{\hbox{$\rm\thinspace ct~keV^{-1}~cm^{-2}~s^{-1}$}}

\def\ergpcmsqps{\hbox{$\rm\thinspace erg~cm^{-2}~s^{-1}$}}

\def\ergps{\hbox{$\rm\thinspace erg~s^{-1}$}}

\def\ug{\hbox{$\rm\thinspace \mu G$}}

\begin{document}

\title[] {The luminous X-ray hotspot in 4C\,74.26: synchrotron or inverse-Compton emission?}
\author[M. C. Erlund et al.]  {\parbox[]{6.in} {M.~C.
    Erlund,$^{1}$\thanks{E-mail: mce@ast.cam.ac.uk} A.~C.
    Fabian,$^{1}$ Katherine~M. Blundell,$^{2}$ C. Moss,$^{3}$\\
    and D. R. Ballantyne$^{4}$}\\\\
  \footnotesize
  $^{1}$Institute of Astronomy, Madingley Road, Cambridge CB3 0HA\\
  $^{2}$University of Oxford, Astrophysics, Keble Road, Oxford OX1 3RH\\
  $^{3}$Astrophysics Research Institute, Liverpool John Moores University, Twelve Quays House, Egerton Wharf, Birkenhead, CH41 1LD\\
  $^{4}$Department of Physics, University of Arizona, 1118 E. 4th Street, Tucson, AZ, U.S.A.}
\maketitle

\begin{abstract}
  We report the discovery of an X-ray counterpart to the southern
  radio hotspot of the largest-known radio quasar 4C\,74.26 (whose
  redshift is $z=0.104$).  Both \xmm\ and \chandra\ images reveal the
  same significant ($10$\as, i.e.  $19$\kpc) offset between the X-ray
  hotspot and the radio hotspot imaged with MERLIN.  The peak of the
  X-ray emission may be due to synchrotron or inverse-Compton
  emission.  If synchrotron emission, the hotspot represents the site
  of particle acceleration and the offset arises from either the jet
  exhibiting Scheuer's `dentist's drill' effect or a fast spine having
  less momentum than the sheath surrounding it, which creates the
  radio hotspot.  If the emission arises from the inverse-Compton
  process, it must be inverse-Compton scattering of the CMB in a
  decelerating relativistic flow, implying that the jet is
  relativistic ($\Gamma\ge 2$) out to a distance of at least 800\,kpc.
  Our analysis, including optical data from the Liverpool Telescope,
  rules out a background AGN for the X-ray emission and confirms its
  nature as a hotspot making it the most-X-ray-luminous hotspot
  detected at low redshift.
  \\
\end{abstract}


\section{Introduction}

4C\,74.26, at redshift $z=0.104$, has the morphology of a typical
\frii\ radio source \citep{FR}, but is on the borderline between \fri\
and \frii\ in terms of its radio luminosity \citep{rileywarner}.  It
has a one-sided jet, which lies at an angle $< 50^{\rm o}$ to the
line-of-sight \citep{pearson92} and may be indicative of beaming
($\geq 0.5c$, \citealt{rileywarner}). The southern radio hotspot on
the jet side is much brighter and more compact than the hotspot
complex on the non-jet side, which is typical of such sources (e.g.
\citealt{laing89}; \citealt{bridle94}).  Optical observations show its
host to be a typical giant elliptical galaxy \citep{riley88}.

Our analysis of X-ray observations of 4C\,74.26 show that the southern
hotspot of 4C\,74.26 is more X-ray luminous than the western hotspot
of Pictor A \citep{pictora} and hotspot D in Cygnus A \citep{cyga}
(see Table \ref{table:lum}). It is at a far greater distance from the
nucleus than either of these two sources making it one of the largest
sources in the local Universe with a projected linear size of
$1.1$\mpc\ (assuming $\rm H_{\rm 0} = 71$\kmpspmpc) \citep{pearson92}.
If the jet is inclined at 45 degrees to the plane of the sky, the
maximum likely angle implied by unification models of quasars and
radio galaxies (e.g.  \citealt{barthel89}; \citealt{antonucci93}),
then the south hotspot is $\sim 800$\kpc\ from the nucleus. A smaller
angle than this canonical maximum would make the deprojected (true)
physical size of this giant radio quasar even more extreme.

The southern hotspot in 4C\,74.26 is also possibly the most spatially
extended radio hotspot yet detected, with the major axis of each
component $\sim 15$\as\ on the sky.  Both the northern and the
southern hotspot complexes are each resolved by the VLA in A
configuration into two components with each member of a pair being of
nearly equal brightness to its neighbour.

Hotspots are thought to occur when particles in a jet pass through a
final shock, in which their bulk kinetic energy is converted to random
energy of the constituent particles \citep{brunettiaccn}.  Evidence
for such re-acceleration comes in the form of hotspot Spectral Energy
Distributions (SEDs), morphologies and polarisation
\citep{brunettiaccn}.  Shock acceleration and radiative loss models
also give good fits to radio--to--optical spectra
\citep{meisenheimer89}.  The low energy cut-off directly detected in
6C\,0905+3955 ($\gamma_{\rm min}\sim 10^3$; \citealt{6c0905}),
assuming it is common, also constrains the particle acceleration
process taking place at the shock.

X-ray emission from hotspots can in principle be produced by a number
of different processes, including synchrotron radiation and
inverse-Compton scattering of a seed photon field, for example the
synchrotron emission produced in the jet itself (known as Synchrotron
Self-Compton, SSC, emission).  The Cosmic Microwave Background (CMB)
also provides a seed photon field, this is likely to be important in
relativistic jets (e.g.  \citealt{tavecchio00}; \citealt{celotti01}).
In a relativistic but gradually decelerating flow, the upstream
electrons can up-scatter the synchrotron emission from the slower
moving downstream regions \citep{georganopouloskazanas03}.

Broad band observations of hotspots have shown that many cannot be
explained by a single synchrotron power law (e.g. \citealt{tavecchio05}).
One-zone synchrotron--inverse-Compton models, where beaming is
ignored, imply that hotspots are far from equipartition or are
composed of extremely small condensations.  That a hotspot is detected
on only one side of a source at high energies in some sources (i.e.
quasars) may be explained by relativistic beaming.  Multi-zone models
such as the possibility of having a fast (highly relativistic) spine
surrounded by a slower layer (e.g.  \citealt{chiaberge00}) and
differential relativistic beaming due to decelerating bulk motions in
the hotspots on the synchrotron and inverse-Compton spectra, in
particular the fast moving flow sees the slow moving downstream flow
which is Doppler boosted \citep{georganopouloskazanas04}, can produce
a gradual increase in the ratio of radio--to--X-ray flux away from the
central source. Both models have been used to explain observations
\citep{tavecchio05}.

Here we present \chandra\ and \xmm\ observations of the southern
hotspot of the powerful, giant, broad-line radio-loud quasar 4C\,74.26
\citep{rileywarner}.  The data used here have already been used in
work on the nucleus by \citet{ballantynefirst} and \citet{ballantyne},
using the \xmm\ data, and by \citet{kaspi04} for the \chandra\ data.
Throughout this paper, all errors are quoted at $1\sigma$ unless
otherwise stated and the cosmology is $\rm H_{\rm 0} = 71$\kmpspmpc,
$\Omega_{0}=1$ and $\Lambda_{0} = 0.73$.  One arcsecond represents
$1.887$\kpc\ on the plane of the sky at the redshift of 4C\,74.26 and
the Galactic absorption along the line-of-sight is $1.19 \times
10^{21}$\pcmsq\ \citep{dickeylockman90}.


\section{Data Reduction}
\label{sec:reduction}

\begin{table} 
\begin{tabular}{llcl}

\hline
Telescope           & Date         & Obs ID        & ksec   \\
\hline
\xmm\               & 2004 Feb. 2  & $0200910201$  & $33.9$\\ 
\chandra\ with HETG & 2003 Oct. 6  & $4000$        & $37.2$  \\
\chandra\ with HETG & 2003 Oct. 8  & $5195$        & $31.4$ \\
\hline  
\end{tabular}
\caption{Details of our X-ray observations: (1) X-ray telescope used 
(2) date of the observation, (3) observation identification number 
and (4) duration of flare-cleaned observations.}
\label{table:obs} 
\end{table}

Table \ref{table:obs} contains a summary of the X-ray observations
analysed. The \xmm\ data was reduced using the Standard Analysis
System (SAS).  Only EPIC-pn data is of interest here as the southern
hotspot is not visible on the MOS chips due to the window used.  The
EPIC-pn observation was taken in Pointing mode (the data mode was
Imaging) with submode PrimeLargeWindow.  The analysis chain EPCHAIN
was run and the data were filtered for periods of heavy flaring. The
astrometry was corrected by aligning the centroid of X-ray nucleus to
the VLA radio core (A-array data), a shift of $1.5$\as.  RMFGEN and
ARFGEN (both SAS tasks) produced the response matrix and ancillary
response files used.  Photons were grouped 20 per bin, using GRPPHA,
for spectral analysis.

The \chandra\ data were processed using the \ciao\ data processing
software package (\ciaov\ and \caldb). Unfortunately, both \chandra\
observations were taken using the HETG which has the effect of greatly
reducing the source counts detected in the zero-order image, but not
the noise. The astrometry was corrected using the Astrometry
Correction
Thread\footnote{http://cxc.harvard.edu/cal/ASPECT/fix\_offset/fix\_offset.cgi}
which corrected \chandra 's positioning by $0.1$\as.  Pixel
randomisation was turned off for all observations using the {\it
  ACIS\_PROCESS\_EVENTS} tool. Then the Sub-pixel Resolution Algorithm
(\citealt{subpix} and \citealt{subpix2}) was used to make use of
photons that arrive near the edges and corners of the pixels, as their
arrival point can be determined with sub-pixel resolution.  The
centroid of the \chandra\ nucleus agrees with the VLA A-array radio
core to $0.1$\as, so there is no need to re-align to the \chandra\
X-ray core with the VLA radio core.  The southern hotspot is
$\sim5$\am\ from the central source and so it was imaged using ACIS-S2
chip (CCD 6). Its \chandra\ and \xmm\ fluxes are consistent within the
errors.

\begin{table*}
\begin{tabular}{ccccc}
  \hline
  Instrument   & date       & frequency    & beam            & RMS  \\
  &            & \ghz\ &   \as$\times$\as\              & $10^{-4}$\jypb   \\
  \hline     
  VLA A   & 1988 Nov. 16 & $1.489$      & $1.7 \times 1.3$    &  $1.35$   \\
  VLA B   & 1989 Apr. 22 & $1.47$       & $5.1 \times 3.7$    &  $0.83$   \\
  MERLIN  & 2006 Jun. 6  & $1.66$       & $0.19 \times 0.15$  &  $1.12$   \\
  \hline
\end{tabular}
\caption{Radio observations: (1) Instrument, with A and B referring to the VLA array configuration, (2) date of the observation, 
(3) frequency of the observations, (4) the beam size (major axis and minor axis each in\as) 
and (5) the root mean-square of the background noise in the observation. }
\label{table:radio} 
\end{table*}

The radio observations were made with the VLA\footnote{The National
  Radio Astronomy Observatory is operated by Associated Universities,
  Inc., under cooperative agreement with the National Science
  Foundation.} and MERLIN.  A summary of the observations can be found
in Table \ref{table:radio}.  The VLA data were processed using
standard techniques within the AIPS package, including
self-calibration for phase only.  The MERLIN data were initially
reduced using the MERLIN pipeline, further processing was done using
standard procedures within AIPS.

Optical data, obtained from the 2 m Liverpool Telescope using the CCD
camera RATCam with an SDSS-r filter (5560 -- 6890 \AA), were taken in
photometric conditions in three consecutive exposures each of 240\s\
on 2006 November 9. Zero-point calibration was provided by
observations of Landolt standard star fields \citep{landolt92} which
bracketed observations of 4C\,74.26 both in time and airmass.  The
uncertainty in the measured zero-point, $\delta m = 0.08$, is
principally due to the error associated with transformation to Johnson
R magnitudes for observations in a single waveband.


\section{Results}

An X-ray source is seen in the \xmm\ image close to the peak of the
southern radio hotspot (Fig. \ref{fig:all}). Its \xmm\ spectrum (Fig.
\ref{fig:spec}) comprises $598$\cts\ (of which $187$\cts\ are due to
background) in the $0.5-10$\kev\ band. The source is detected in the
\chandra\ observations along with some diffuse emission to the south
which is not resolved by \xmm.  Unfortunately, there are not enough
photons to constrain even a crude spectrum because both observations
were taken with the HETG grating in place (summing the two
observations which have roughly equal numbers of counts, there are a
total of $53$ counts, of which $9$ are from the background, in the point
source and a further $35$ counts, with $13$ background counts, in the
diffuse emission).

Fitting a Galactic absorbed power-law to the \xmm\ data gives a photon
index of $\Gamma = 1.54\pm 0.10$ with a reduced chi-squared of
$\chi^{2}_{\nu} = 0.93$ for 18 degrees of freedom and an absorption
corrected X-ray flux of $2.6 \times 10^{-14}$\ergpcmsqps\ in the
$0.5-2$\kev\ band and $5.9\times 10^{-14}$\ergpcmsqps\ in the
$2-10$\kev\ band.  The absorption corrected X-ray luminosity in the
$0.5-10$\kev\ is $2\times 10^{42}$\ergps, assuming it is at the
redshift of 4C\,74.26.  Table \ref{table:results} shows the results of
the spectral fitting.

The brightest parts of the southern radio hotspot are resolved, with
MERLIN's $0.2$\as\ resolution, into a complicated hotspot complex (see
Fig.  \ref{fig:allradio}). The MERLIN hotspot consists of two peaks,
both of which are resolved, with wings of radio emission attached.
The east and west peaks seem to be linked by a bridge of diffuse
emission, where the wings of the two peaks almost meet.  The
westernmost peak and wings are the brightest component, the
easternmost wing of this hotspot component ends with a bright arc.

The peak of the X-ray emission is offset from the brightest radio
emission in the southern hotspot complex (the westernmost, brightest
peak in the MERLIN data) by $\sim 10$\as. (It is offset from the
brightest peak in the VLA-array data by $6$\as.) This offset is
clearly illustrated in Fig. \ref{fig:profile}, as well as in Fig.
\ref{fig:all}, Fig.  \ref{fig:xray}, and Fig.  \ref{fig:allradio}.
The position of the X-ray source detected with both \xmm\ and
\chandra\ coincides to $< 0.6$\as\ (see Fig.  \ref{fig:xray}) making
it unlikely that the offset is due to instrumental effects which would
be telescope dependent.  The positional uncertainty is $<2$\as\ in the
\xmm\ data (\citealt{xmmpositioning} report that the $1\sigma$ rms
absolute astrometric accuracy\footnote{see \xmm\ calibration document
  Kirsh et al.  XMM-SOC-CAL-TN-0018 at http://xmm.vilspa.esa.es.} is
about 2\as, we note that this relates to raw \xmm\ data and so our
radio-aligned \xmm\ observation will have more accurate astrometry
consistent with the relative astrometry for the EPIC-pn, which is
$1.5$\as, and the accuracy of the position of the VLA radio core).
The positional uncertainty in the \chandra\ data is $\sim 0.6$\as\ at
$5$\am\ off-axis \citep{positionaluncertainty}, $1.7$\as\ for the
radio A-array data, $5$\as\ for the B-array data, and $0.2$\as\ for
the MERLIN data.  This offset is therefore real.

The peak of the MERLIN emission is not consistent with the peak of the
VLA A-array emission.  They are offset by $\sim 4$\as\ (Fig.
\ref{fig:profile}).  Measuring the flux density in the MERLIN hotspot
complex (see Fig. \ref{fig:merlinopt}) and comparing it to the flux
density in the VLA A-array hotspot, we note that they differ by a
factor of $\sim 2$ ($0.23$\jy\ for MERLIN at $1.66$\ghz\ and
$0.47$\jy\ for VLA A-array at $1.49$\ghz).  The difference between the
MERLIN and the VLA flux densities tells us that there is significant
extended structure associated with the hotspot complex which the
shorter baseline of the VLA is sensitive to but which MERLIN does not
detect; MERLIN only detects the compact structure.

The northern hotspot complex, which is much weaker in the radio, is
$\sim 5$\am\ north of the core and is not detected either in the
\chandra\ data (it would lie on the edge of ACIS-S3 chip) or the \xmm\
data.  The upper limit of its X-ray flux, calculated from the \xmm\
data, is $3.3\times 10^{-15}$\ergpcmsqps\ in the $0.5-10$\kev\ band.
So the ratio of hotspot north--to--south X-ray fluxes is $\ge 22$,
assuming that the X-ray source is the hotspot.

The radio energy flux (in $\nu F_{\nu}$ -- i.e. the flux density
multiplied by the frequency of the observation) of the southern and
northern radio hotspot complexes, using B-array data and considering
the whole hotspot region not just the area coincident with the X-rays,
is $7.4\pm 0.1\times 10^{8}$\jy\hz\ and $6.3\pm 0.1 \times
10^{7}$\jy\hz\ respectively.  The X-ray energy flux at $1$\kev\
(calculated by converting the X-ray power-law normalisation which is
in \photpkevpcmsqps\ at $1$\kev\ to \jy\hz, using $1$\photpkevpcmsqps\
$= 1.6\times 10^{14}$\jy\hz\ and the normalisation given in Table
\ref{table:results}) is $1.8\pm 0.2\times 10^{9}$\jy\hz\ for the
southern hotspot and $<1.3\times 10^8$\jy\hz\ for the upper-limit on
the northern hotspot.  The southern hotspot emits roughly twice as
much energy in X-rays as in radio emission. The northern hotspot
X-ray--to--radio flux ratio is $\ll 2$.

No counterpart to the X-ray source was found in the DSS
images\footnote{http://www-gsss.stsci.edu/SkySurveys/Surveys.htm} or
those from 2MASS and \xmm\ Optical Monitor (there are no available
\hst\ images).  Our Liverpool Telescope observations go much deeper
($R = 24.3$ at $3\sigma$ for a point source); however, no optical
source at the position of the X-ray source is detected, but there is
some faint diffuse emission consistent with being coincident with the
westernmost peak of the radio hotspot as resolved by MERLIN (see Fig.
\ref{fig:merlinopt}) which has a magnitude of $R=22.2\pm 0.9$ (using a
$7$\as\ aperture), and is detected at the $2.4\sigma$ level (see Fig.
\ref{fig:optical}).  Deep high-resolution multi-band optical
observations are required to confirm the presence of this optical
hotspot, to measure its flux with much more certainty, and especially
to resolve and characterise it.

The optical R band energy flux, which is calculated by multiplying the
flux density\footnote{Calculated using
  http://www.astro.soton.ac.uk/\~{}rih/applets/MagCalc.html} by the
observation frequency $4.28\times 10^{14}$\hz\ (i.e. at a wavelength
of 7000\a), is $1.7\pm 0.9 \times 10^9$\jy\hz\ for the diffuse
emission coincident with the radio hotspot and $< 2.4 \times
10^8$\jy\hz\ for the X-ray hotspot. The energy radiated in X-rays,
from the X-ray hotspot, and in optical R band emission, coincident
with the radio hotspot, is consistent within the errors; although it
should be remembered that the optical detection is only at $2.4\sigma$
so there is considerable uncertainty in this measurement.

The jet is not detected by \chandra\ and a scattering spike on the
\xmm\ image exactly coincides with the radio jet direction, making any
quantitative constraint on X-ray emission from the jet impossible.  We
note that the second lowest contour in Fig. \ref{fig:all} extends
further in the jet direction than on the other spikes so jet emission
at that level is likely.

\subsection{X-ray hotspot or coincidental source?}

\begin{figure}
\rotatebox{0}{
\resizebox{!}{13cm}
{\includegraphics{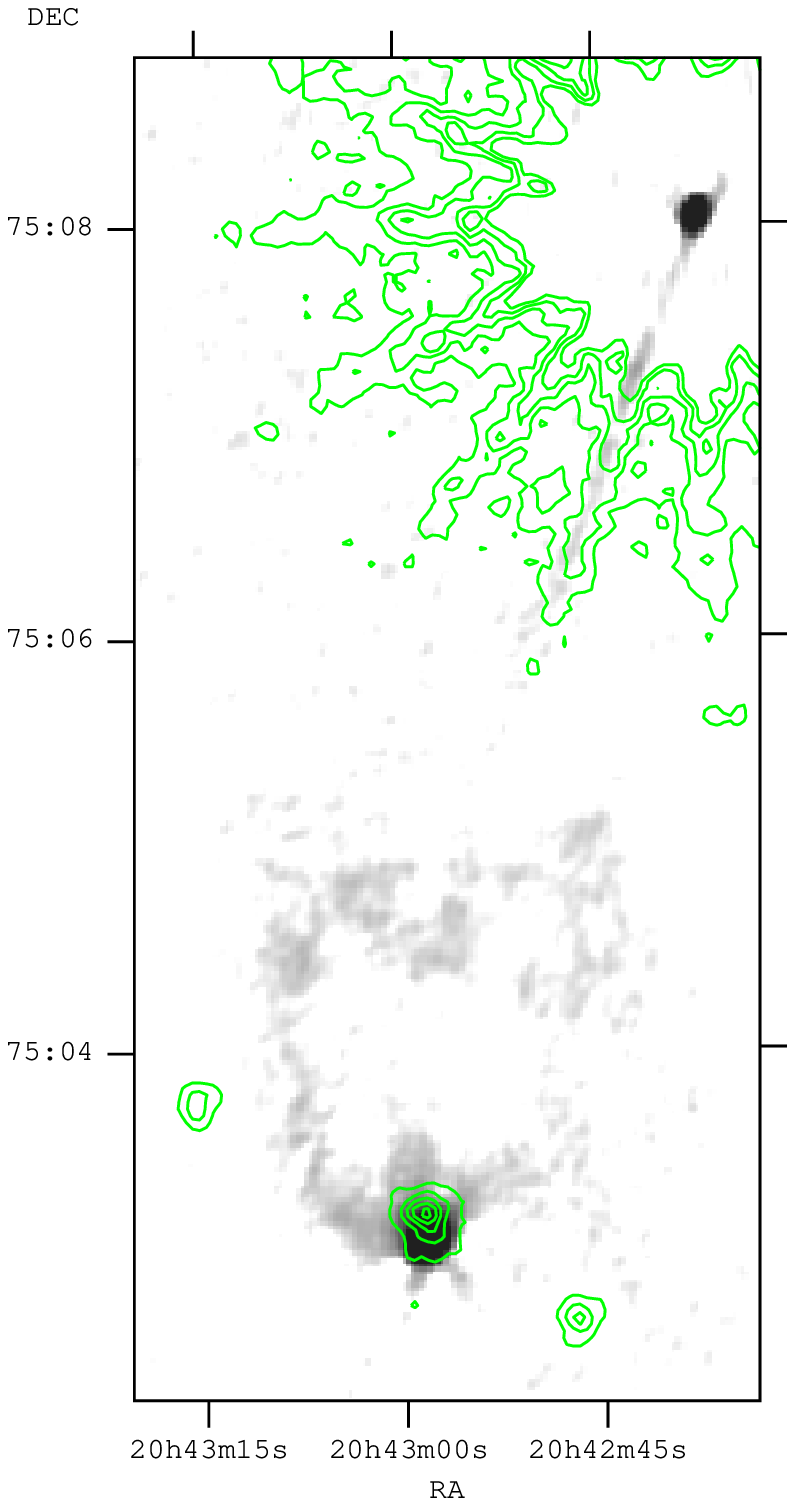}}
}
\caption{Radio grey-scale (the beam is $5.1$\as\ by $3.7$\as\ using
  the B-array), the green contours are from the \xmm\ $0.5-10$\kev\
  band image, smoothed by $2$\as\ ($1.4$, $2.8$, $4.2$, $5.6$ and
  $7$\cts\ per $2$\as\ pixel).  The X-ray sources visible to the east
  and west both have easily identifiable optical counterparts with R
  magnitudes of $17.4$ and $19.1$ respectively.}
\label{fig:all}
\end{figure}

\begin{figure}
\rotatebox{270}{
\resizebox{!}{9cm}
{\includegraphics{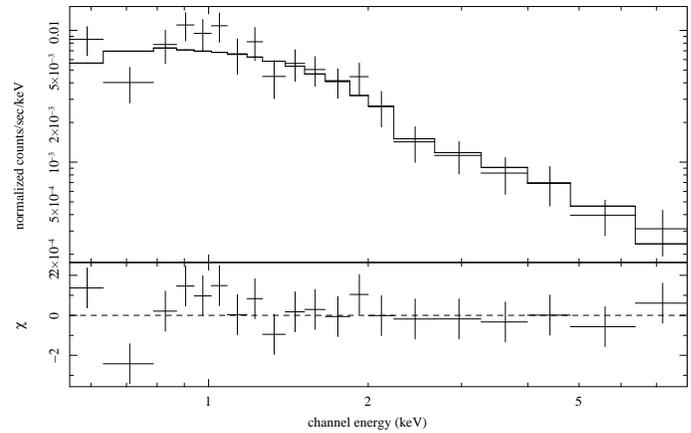}}
}
\caption{\xmm\ spectrum in the $0.5-10$\kev\ band with $20$ photons
  per bin.  The solid line represents the best fit power-law with
  Galactic absorption fixed at $1.19\times 10^{21}$\pcmsq\ and a
  photon index of $\Gamma=1.54\pm 0.17$.  For more details see Table
  \ref{table:results}.}
\label{fig:spec}
\end{figure}


\begin{figure}
\rotatebox{0}{
\resizebox{!}{6.3cm}
{\includegraphics{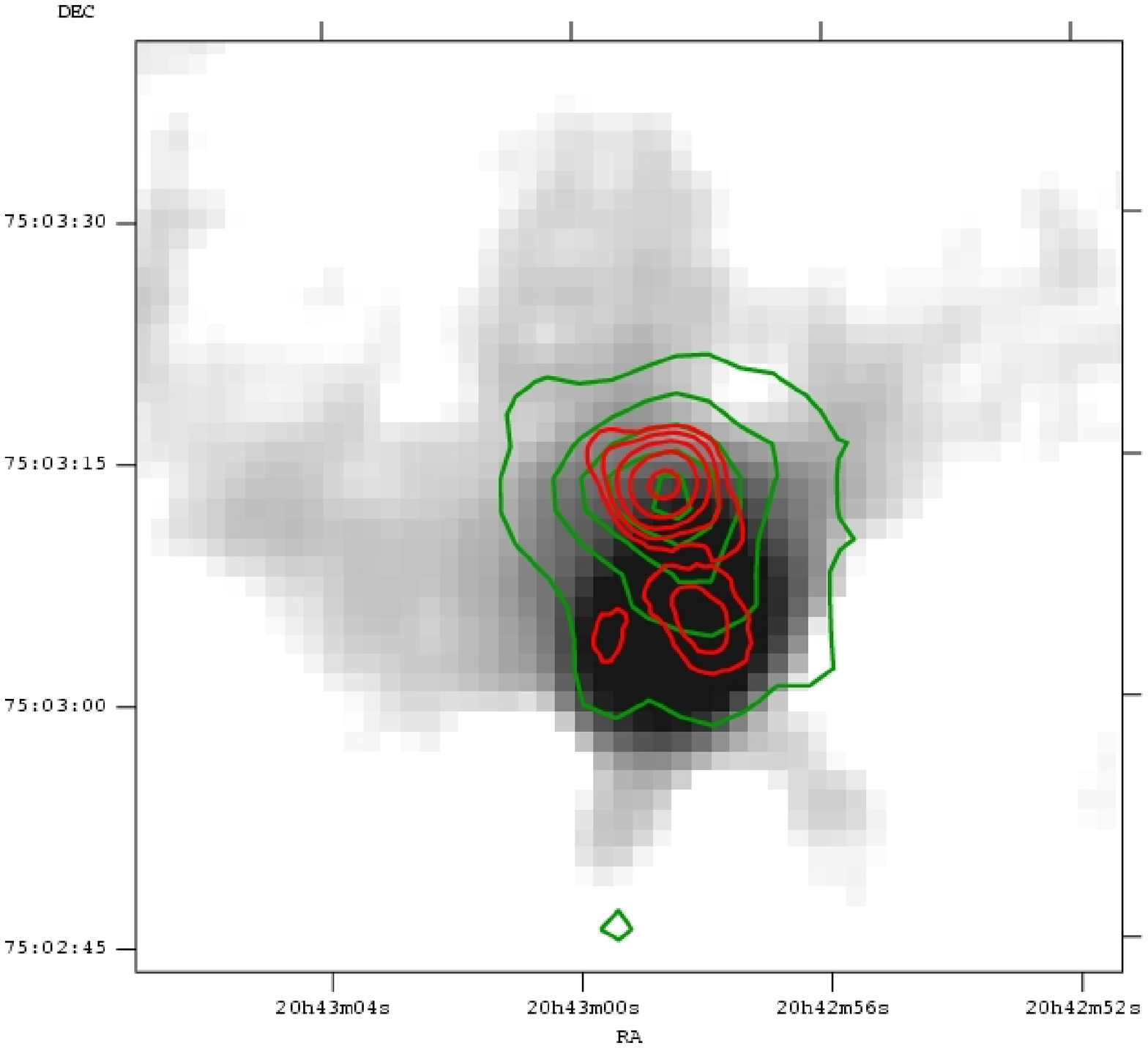}}
}
\caption{$1.47$\ghz\ radio data (the beam is $5.1$\as\ by $3.7$\as\
  using the B-array) is shown in grey-scale with X-ray contours
  overlaid.  The green contours are taken from the \xmm\ $0.5-10$\kev
  -band image smoothed by $2$\as\ and the contour levels are $1.4$,
  $2.8$, $4.2$, $5.6$ and $7$\cts\ per $2$\as\ pixel. The red contours
  are from the stacked \chandra\ $0.5-7$\kev\ images smoothed by
  $1.5$\as\ with contour levels $0.07$, $0.11$, $0.17$, $0.26$ and
  $0.40$\cts\ per $0.49$\as\ pixel. \xmm\ centroid is at 20h42m58.7s
  +75d03m13s, the \chandra\ centroid is at 20h42m58.6s +75d03m13s.}
\label{fig:xray}
\end{figure}

\begin{figure}
\centering
\rotatebox{0}{
\resizebox{!}{8cm}
{\includegraphics{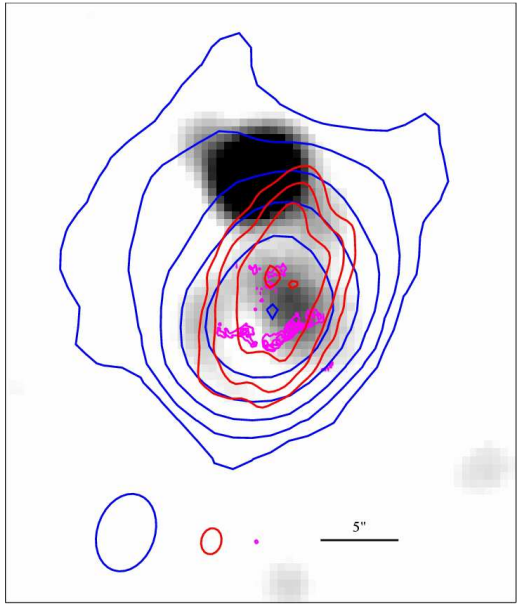}}
}
\caption{Chandra data shown in grey-scale ($0.5-7$\kev -band,
  $0.49$\as\ pixels smoothed by $3$ pixels) overlaid with blue
  contours from the VLA B-array (at $1.47$\ghz; contour levels: $1.0$,
  $2.9$, $8.3$, $24$, $69$, $200$\mjy); red contours from the VLA
  A-array (at $1.49$\ghz; contour levels: $2.0$, $4.3$, $9.3$,
  $20$\mjy) and magenta contours from MERLIN (at $1.66$\ghz; contour
  levels: $0.5$, $1$, $1.5$, $2.0$\mjy).  The beam sizes and
  orientations of the various observations are shown in the same
  colour as their respective contours. }
\label{fig:allradio}
\end{figure}

\begin{figure}
\centering
\rotatebox{0}{
\resizebox{!}{8cm}
{\includegraphics{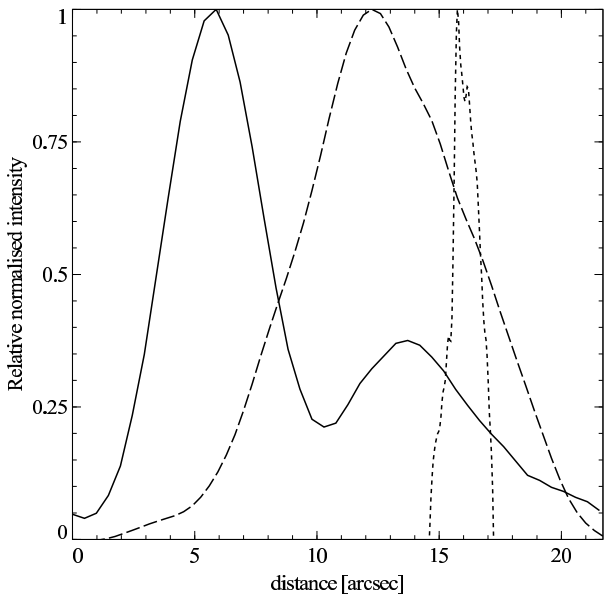}}
}
\caption{North-south profile (left is north, right is south) through
  the X-ray--Radio hotspot complex, using a box $5.4$\as\ wide.  The
  solid line is the \chandra\ data (smoothed by 1.5\,arcsec), the
  dashed line is the VLA A-array data and the peak of the MERLIN data
  is shown by the dotted line. All profiles have been normalised to 1
  at the peaks. }
\label{fig:profile}
\end{figure}

\begin{figure*}
\rotatebox{0}{
\resizebox{!}{8cm}
{\includegraphics{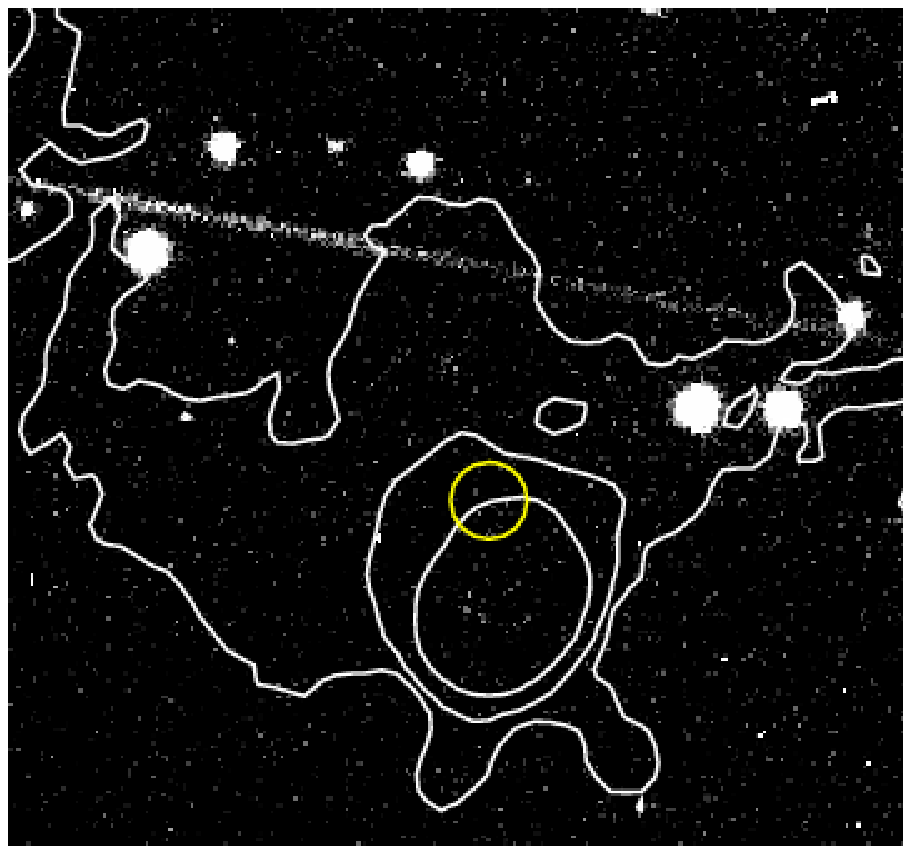}}
}
\rotatebox{0}{
\resizebox{!}{8cm}
{\includegraphics{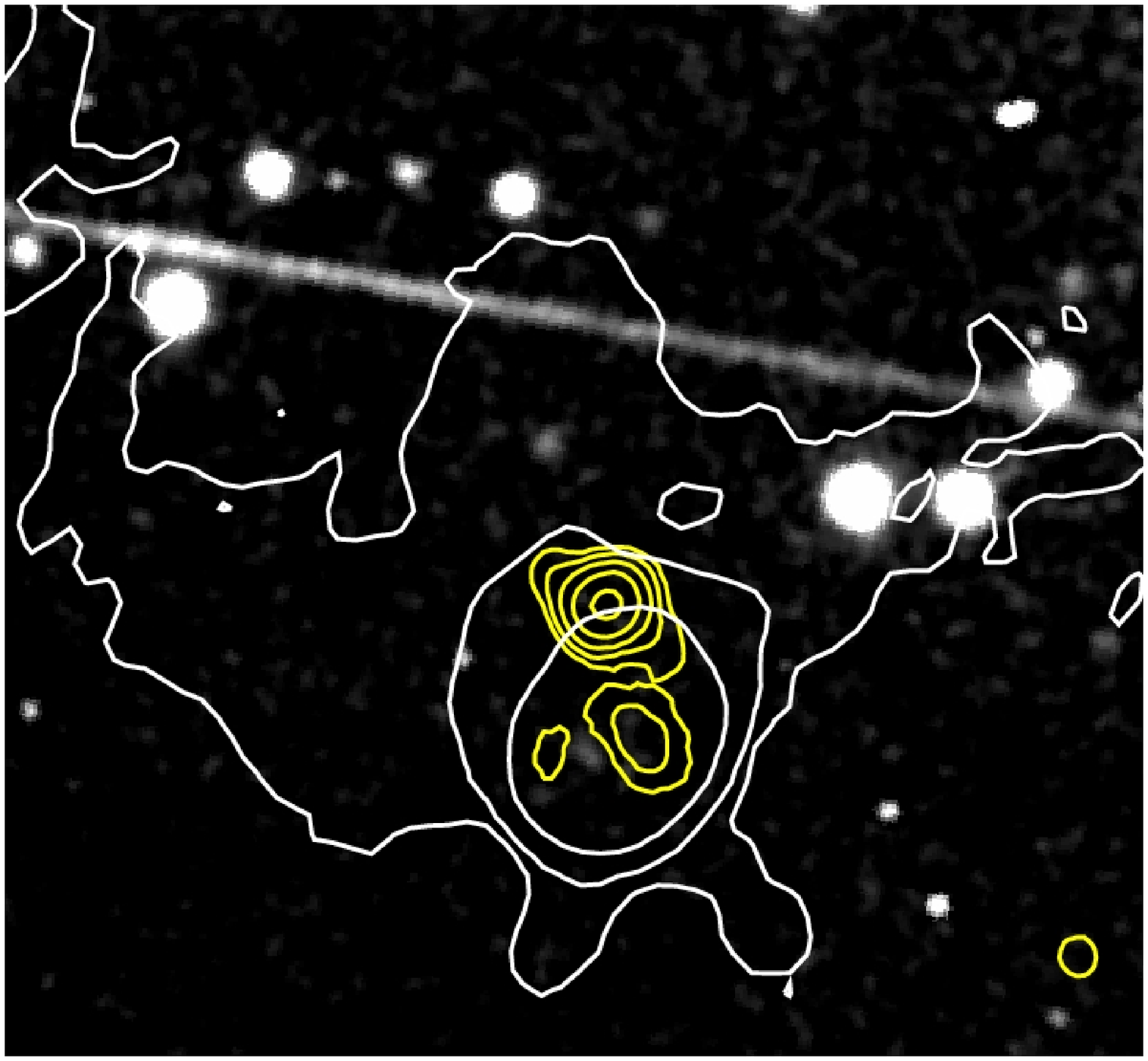}}
}
\caption{Optical image from the Liverpool telescope overlaid with a
  yellow circle (3\as\ radius) at the location of the X-ray source
  (using the position from \xmm\ data) and white contours from the VLA
  B-array data (contours at $3.0$, $1.7$ and $10$\mjy). {\it Left:}
  unsmoothed optical image ($0.28$\as\ pixels). {\it Right:} optical
  data Gaussian-smoothed by $0.84$\as. The yellow contours are from
  the stacked \chandra\ $0.5-7$\kev\ images smoothed by $1.5$\as\ with
  contour levels $0.07$, $0.11$, $0.17$, $0.26$ and $0.40$\cts\ per
  $0.49$\as\ pixel. The almost horizontal white streak is due to a
  persistence effect from a bright star on the immediately preceding
  frame which was for another program; however, it does not affect our
  results. }
\label{fig:optical}
\end{figure*}

To check whether the X-ray source is consistent with being a
background AGN, we calculate the X-ray--to--optical flux ratio, $X/O$,
which is $> 120$, using $\log X/O = \log F_{\rm X} + 0.4 R + 5.6$
\citep{giovanniULX}, where $F_{\rm X} = 5.9\times
10^{-14}$\ergpcmsqps\ is the X-ray flux in the $2-10$\kev\ band and
$R$ is the $3\sigma$ R band magnitude limit.

The cores of typical AGN have $X/O$ ratios of $<10$ and most AGN with
$X/O>10$ are heavily obscured, Type II AGN \citep{mignoli04}. Sources
with $X/O>100$ are very rare and are known as Extreme X-ray / Optical
sources (EXO) \citep{koekemoer04}. If this source were an EXO then it
would be the most X-ray bright example detected to date, which makes
estimating the probability of finding such an EXO in the field of
4C\,74.26 very difficult (see Fig. 2 in \citealt{CDFN}, Table 1 in
\citealt{mignoli04}, as well as \citealt{koekemoer04},
\citealt{grothstrip} and \citealt{CDFS}).  The X-ray spectra do not
require excess absorption, so we conclude that the source cannot be an
obscured AGN.

The \chandra\ observations show diffuse emission to the south of the
X-ray hotspot peak.  At $5$\am\ off-axis the \chandra\ PSF is such
that it is currently impossible to say whether this peak is a point
source or an extended one.  \xmm's PSF is so broad that the diffuse
emission clearly visible in the \chandra\ data is only marginally
detected with \xmm.

\begin{figure}
 \centering
\rotatebox{0}{ 
 \resizebox{!}{8cm}
 {\includegraphics{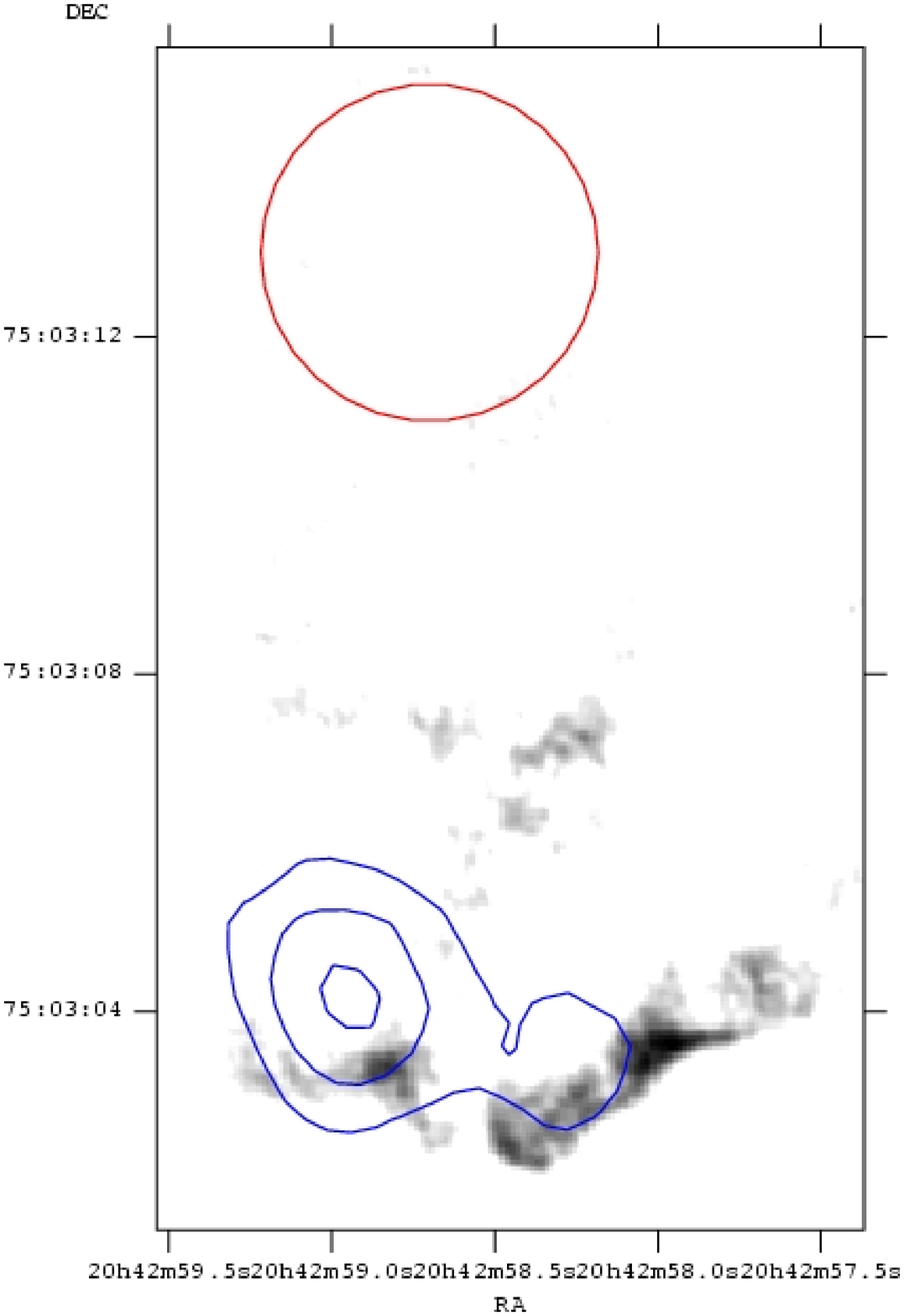}} }
\caption{ 1.66\,GHz radio image in grey-scale from MERLIN overlaid
  with blue contours from SDSS-r band optical data (smoothed by
  0.84\,arcsec, from the Liverpool Telescope) and with a red circle at
  the location of the X-ray hotspot (from XMM-Newton data).}
\label{fig:merlinopt}
\end{figure}

\begin{table*} 
\begin{tabular}{lcccc}
  \hline
  $N_{\rm H}$             & $\Gamma$               & Normalisation                   & $\chi^{2}_{\nu}$ (dof) & $L_{\rm X}$  \\
  $[\times 10^{21}$\pcmsq\ $]$  & & $[\times 10^{-5}$\photpkevpcmsqps\ at $1$\kev\ $]$ &             & $[\times 10^{42}$ \ergps\ $]$ \\
  \hline
  $1.19$                 & $1.54^{+0.10}_{-0.10}$ & $1.13^{+0.09}_{-0.09}$ & $0.93$ ($18$)          & $2.2^{+0.3}_{-0.3}$       \\
  $1.21^{+0.06}_{-0.03}$ & $1.53^{+0.11}_{-0.38}$ & $1.12^{+0.34}_{-0.25}$ & $0.99$ ($17$)          & $2.2^{+0.7}_{-0.7}$       \\
  \hline     
\end{tabular}
\caption{\xmm\ spectrum in the $0.5-10$\kev\ band fitted with an absorbed power-law.  
  In the first line, the absorption has been fixed to the Galactic value in the direction 
  of 4C\,74.26 and in the second line it has been left as a free parameter.  $L_{\rm X}$ 
  is measured in the $0.5-10$\kev -band and has been corrected for absorption.}
\label{table:results} 
\end{table*}

\begin{table} 
  \begin{tabular}{ccc}
\hline
source       & $z$               & $L_{\rm HS}(0.5-10)$\kev\ \\
             &                   & $[10^{42}$\ergps $]$   \\
\hline     
Pictor A     & $0.035$           & $1.6$                  \\
Cygnus A     & $0.0565$          & $1.8$                  \\
4C\,74.26    & $0.104$           & $2.2$                  \\
\hline
\end{tabular}
\caption{The luminosity of the brightest X-ray hotspots, calculated using the power-law normalisation and photon 
  indexes as found in \citet{pictora} and \citet{cyga} for Pictor A and Cygnus A, respectively, and this paper.}
\label{table:lum} 
\end{table}

Typical values for hotspot photon indices found in the literature
range from $\Gamma = 1.4 \pm 0.2$ for hotspot A$_2$ in 3C\,303
\citep{kataoka03} and $\Gamma = 1.5\pm 0.1$ for northern hotspot J in
3C\,351 \citep{hardcastle02} to $\Gamma = 1.0\pm 0.3$ for southern
hotspot K in 3C\,263 \citep{hardcastle02}.  The southern hotspot, for
which the photon index is $\Gamma = 1.54\pm 0.10$, is therefore
consistent with the reported range of X-ray hotspot photon indices.


The preceding analysis shows that it is most likely that the X-ray
source is an X-ray hotspot offset from the radio peak but still lying
just within the region covered by the extended radio hotspot (see VLA
A- and B-array contours in Fig. \ref{fig:allradio}), to the north of
its peak (Fig. \ref{fig:merlinopt}) and along the axis of the radio jet
as it enters the radio hotspot (Fig. \ref{fig:all} and Fig.
\ref{fig:xray}). In Fig. \ref{fig:allradio}, it can be clearly seen
that the X-ray peak is offset from the westernmost peak and brightest
part of the complicated radio hotspot.  There is some faint X-ray
emission associated with the radio hotspots themselves. A \chandra\
observation (without the gratings in place) of the X-ray hotspot
complex would enable us to characterise the peak and diffuse emission
both spatially and spectrally, which is essential for understanding
the processes taking place in this source.

\section{discussion}
\label{sec:discus}

The southern X-ray hotspot in 4C\,74.26 is unusual, not only because
it is the most luminous X-ray hotspot to be detected at low redshift
(see Table \ref{table:lum}) and probably furthest from its central
nucleus, but also because of the significant offset between its X-ray
and radio hotspots (Fig. \ref{fig:profile}, $10$\as\ using MERLIN
data, which is $\sim 19$\kpc\ on the plane of the sky at the redshift
of 4C\,74.26).

4C\,74.26 is not the first nearby source to be detected with an offset
between its X-ray and radio hotspot components. 4C\,19.44 ($z=0.72$)
has a $2$\as\ ($14.5$\kpc) offset between its X-ray and radio hotspots
which is in the same sense as the offset in 4C\,74.26---it is also a
large ($195$\kpc\ jet), one-sided source \citep{sambruna02}.

Offsets have been detected between an increasing number of radio,
optical and X-ray jet-knots in large-scale jets.  The offset is
usually in the same sense as for 4C\,74.26 with the peak of the X-ray
emission lying closer to the nucleus than the peak of the radio
emission (e.g.  PKS\,1127-145, \citealt{siemiginowska02}; M87,
\citealt{marshall02}; 3C\,273, \citealt{jester06}).

\subsection{What is the X-ray emission mechanism? }

In both the northern and southern radio hotspot complexes, the
multiple components are close together, arguing against either
component being a splatter spot in the manner suggested by the
simulations of \citet{coxgullscheuer91}.  Multi-frequency radio
imaging will show whether the hotspots have different spectra and
hence indicate whether one is consistent with being the ``primary''
hotspot.

X-rays from a bow shock ahead of the radio hotspot can be ruled out as
the X-ray emission is behind, not in front, of the radio hotspot.
X-ray emission associated with hotspots is only likely to be
synchrotron, inverse-Compton or synchrotron self-Compton (SSC).  The
latter can be ruled out for the X-ray peak precisely because the X-ray
and radio emission are not co-spatial.  Inverse-Compton, SSC or
synchrotron emission may be responsible for the diffuse X-ray emission
associated with the radio hotspots.

\subsubsection{Inverse-Compton}

X-ray emission from inverse-Compton scattering of CMB photons requires
electrons with Lorentz factors, $\gamma$, of only $\sim 10^3$, much
lower than the likely Lorentz factors of synchrotron radio-emitting
electrons.  This can lead to a separation in X-ray and radio peaks if
there is a low-energy cutoff to the electron spectrum at about
$\gamma\sim 10^3$ so that X-rays only arise when the population has
cooled sufficiently, and so trail behind the radio emission region
\citep{6c0905}.  The compact nature of the X-ray hotspot makes this
explanation unlikely in this case.

\begin{figure}
\rotatebox{-90}{
\resizebox{!}{8cm}
{\includegraphics{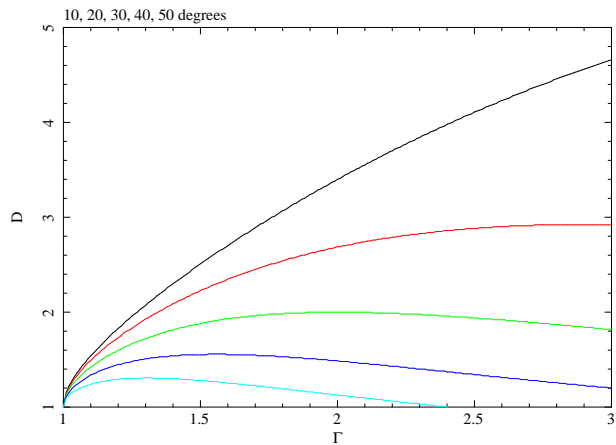}}
}
\caption{Doppler beaming parameter ($D$) as a function of the bulk
  Lorentz factor of the jet ($\Gamma$).  The different lines represent
  different angles varying from 10$^\circ$ to 50$^\circ$ (from top to
  bottom). }
\label{fig:eccmb}
\end{figure}

\citet{georganopouloskazanas04} showed that gradual deceleration of a
jet after a relativistic shock can produce a separation of X-ray and
radio peaks.  The X-ray emission is presumed due to inverse-Compton
scattering of relativistic electrons on the CMB.  In order for the
X-ray peak to be significantly displaced from the end of the hotspot
(where the radio peaks), the angle of the jet to the line-of-sight
must be less than $\sim 40^\circ$, and preferably less than
$30^\circ$, and the bulk Lorentz factor of the jet at the X-ray spot
$\Gamma < 2$ (see Fig. \ref{fig:eccmb}). The sheer size of 4C\,74.26
means that if it were to be aligned closer to the line-of-sight than
$14^\circ$, it would have a deprojected size larger than that of the
largest known radio galaxy 3C\,236 which spans $\sim 4.4$\mpc\
\citep{macketal97}.  The beamed emission varies as a function of the
Doppler beaming parameter, $D$ (see Fig. \ref{fig:eccmb}), to a high
power ($\sim 5$), which means the X-ray emission can drop sharply from
the peak to where the jet stops, which is where the radio emission
peaks due to compression of the magnetic field as the jet slows down.
The synchrotron (radio) emission also has an intrinsically lower
beaming dependence. It should be noted that this model relies heavily
on beaming, so if the jet was aligned closer to our line-of-sight the
X-ray hotspot would be very much brighter. The X-ray hotspot in
4C\,74.26 is already the most X-ray luminous in the local Universe
($z<0.2$).

\subsubsection{Synchrotron}

We now discuss the feasibility of the X-ray hotspot being synchrotron
in origin.  If this is the case, then the X-ray emission represents
the main site of the particle acceleration, since the synchrotron
radiative lifetime at X-ray energies is very short.  The synchrotron
cooling time of X-ray emitting electrons in the $10-100$\ug\ fields
typical of hotspots is $30$ to $10^3$\yr\ whereas for radio emitting
ones it is $4\times 10^5$ to $10^7$\yr.

The observed X-ray photon index, $\Gamma = 1.54\pm 0.10$ (i.e. $\alpha
= 0.54\pm 0.10$) is consistent with theoretical predictions for shock
acceleration models (\citealt{heavensdrury88},
\citealt{achterberg01}).  The X-ray hotspot is more luminous than the
radio hotspot, consistent with the idea that it is the location at
which most of the jet's bulk energy is converted into the random
motion of its particles.

\begin{figure}
\rotatebox{0}{
\resizebox{!}{8cm}
{\includegraphics{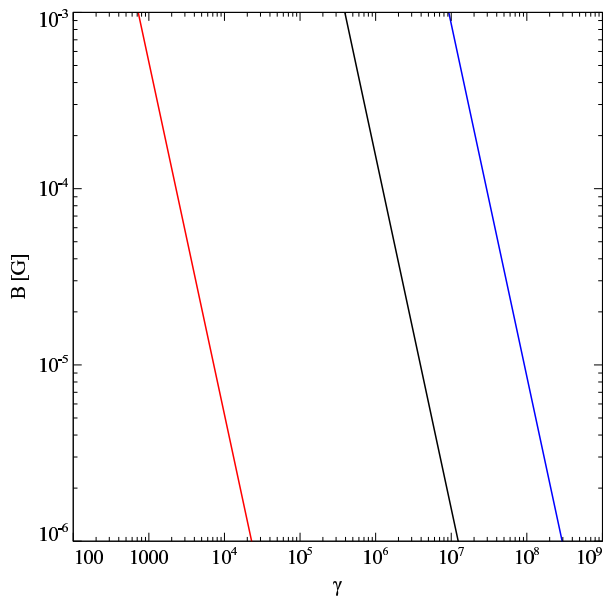}}
}
\caption{Figure showing the relationship between the magnetic field
  strength, $B$, and Lorentz factor of an electron, $\gamma$, (from
  Equation \ref{eqn:gB}) which produces radio synchrotron emission at
  1.47\ghz\ (red line), optical emission at 7000\a\ (black line), or
  X-ray emission at 1\kev\ (blue line).}
\label{fig:gB}
\end{figure}

The location of the shock where particles are accelerated is where the
highest Lorentz factors occur, the magnetic field is compressed and
synchrotron emission is expected to be strongest.  Both high Lorentz
factors and a strong magnetic field contribute to synchrotron emission
being observed at X-ray frequencies, since
\begin{equation}
\nu = \frac{e B \gamma^2 }{2 \pi m_{\rm e}},
\label{eqn:gB}
\end{equation}
where $\nu$ is the frequency of the synchrotron emission, $B$ is the
magnetic field strength, $e$ is the charge on an electron, $m_{\rm e}$
is the rest mass of an electron and $\gamma$ is the Lorentz factor of
the electron. Figure \ref{fig:gB} shows this relationship between
$\gamma$ and $B$, for radio synchrotron (red line), for optical
synchrotron (black line) and for X-ray synchrotron (blue line),
illustrating that in a typical hotspot environment the production of
shorter wavelength radiation will be easier.

The synchrotron cooling model of \citet{bailee03} predicts an X-ray /
radio offset in the sense that we observe in 4C\,74.26's hotspots for
knots along jets.  However, this synchrotron cooling picture alone is
insufficient to explain the nature of the X-ray / radio hotspot offset
because the radio luminosity, while significantly lower than the X-ray
luminosity, is several orders of magnitude higher than predicted by
their model.

A second toy synchrotron model requiring some fine-tuning is one where
the jet possesses a fast spine with lower momentum than the
surrounding slower sheath (e.g. \citealt{chiaberge00}).  This would
mean that the spine would be stopped by the surrounding medium more
easily.  As the spine is moving faster, the shock is more efficient at
accelerating the particles, producing X-ray synchrotron emission,
whereas downstream the sheath, which is moving slower, is less
effective from this point of view, so we see much less X-ray emission
(assuming higher bulk Lorentz factors lead to higher $\gamma$
electrons being produced).

The most likely synchrotron model for the X-ray--radio hotspot offset
is the `dentist's drill' model \citep{scheuer82} where the X-ray
hotspot is the current active end of the jet and the radio hotspot
marks where it was previously.  Presumably the direction of the jet
has moved, either as a result of precession or a perturbation in the
nucleus, or of buoyancy effects or instabilities along the jet.  Note
that the jet must curve, because it clearly does not lie on a straight
line from the nucleus to any hotspot component (Fig.  \ref{fig:all}).
Multiple hotspots, as indeed we have in this object, have previously
been accounted for by such a `dentist's drill' effect
\citep{scheuer82}.  The X-ray hotspot is more luminous than the radio
one which accords with it being the most recent one, in a region of
enhanced magnetic field and synchrotron lifetime arguments fit with it
being young ($30$ to $1000$\yr).  The X-ray hotspot being inset from
the radio hotspot is also consistent with this model as the jet may
now be hitting the side of the cavity it carved out when creating the
radio hotspot. In the context of this model, giant radio sources are
thought to be those that do not move around very much so have created
a channel a long way from their central sources \citep{scheuer82},
this is consistent with the compact double radio hotspot in 4C\,74.26.
It may be that we are seeing the `dentist's drill' effect as an offset
in this source precisely because it is large; 4C\,19.44 is also a
large source where an analogous offset is seen \citep{sambruna02}.

\subsubsection{Observational tests}

The `dentist's drill' model predicts optical/IR emission coincident
with the X-ray hotspot and/or at the location of the no-longer-fed,
radio hotspots.

We detect faint diffuse optical emission associated with the radio
hotspot which is in keeping with our synchrotron models.  Deeper
optical or IR observations are needed to be able to make a more secure
detection and identify its nature.

The inverse-Compton up-scattering of the CMB in a decelerating jet
model predicts a smooth change in X-ray brightness between the X-ray
and radio peak and no change to the spectral index.

Further \chandra\ observations (without the gratings in place) would
permit us to characterise the X-ray hotspot both spatially and
spectrally and thus say whether we are seeing a jet which is
relativistic on scales of at least 800\kpc.

\section{conclusions}
\label{sec:conclude}

We have presented archival \xmm\ and \chandra\ observations of
4C\,74.26, one of the largest sources in the local Universe with each
jet $570$\kpc\ long projected onto the plane of the sky making it the
largest known radio quasar.  We find a bright X-ray counterpart to the
southern radio hotspot which is the most X-ray luminous hotspot known
at low redshift.

There is a significant $10$\as\ offset between the radio (MERLIN) and
X-ray (\xmm) peak; however, our analysis shows that the X-ray source
is not a background AGN due to its extreme X-ray--to--optical flux
ratio and spectrum.

The X-ray hotspot, being more luminous than the radio hotspot, may
reveal the site where the jet dumps most of its energy and is the site
of particle acceleration; the separation of the radio and X-ray
hotspots may arise from Scheuer's `dentist's drill' effect.  The
offset may alternatively be created by inverse-Compton scattering of
the CMB in a relativistic and decelerating jet, which would imply that
the jets in 4C\,74.26 are relativistic on scales of at least 800\kpc.
A final toy model invokes the idea that the jet may contain a fast
spine which has less momentum than its surrounding sheath and is
stopped more efficiently.

In order to differentiate between these various models whose
implications all have important consequences for our understanding of
dynamics of jets, we require higher resolution X-ray data and deeper,
high resolution, optical imaging data.

\section*{Acknowledgements}

MCE acknowledges PPARC for financial support. ACF and KMB thank the
Royal Society. CM thanks LJMU for support and award of LT observing
time. DRB is supported by the University of Arizona Theoretical
Astrophysics Program Prize Postdoctoral Fellowship. We thank Tom
Muxlow for help with the MERLIN observation. We thank Elena Belsole,
Annalisa Celotti and Carolin Crawford for their helpful comments. We
thank the referee for useful comments that improved the content of
this paper. MERLIN is a National Facility operated by the University
of Manchester at Jodrell Bank Observatory on behalf of PPARC.

\bibliographystyle{mnras} 
\bibliography{mn-jour,hotspots.bib}
\end{document}